\documentclass[a4paper]{jpconf}
\usepackage{graphicx}
\usepackage{array}
\usepackage{url}
\usepackage{float} 
\usepackage{xcolor}
\usepackage{wrapfig} 

\usepackage{amsmath,amssymb,graphicx}
\usepackage{hyperref}

\usepackage{listings}
\begin{document}
\title{Exploring Large Language Models (LLMs) through interactive Python activities}
\author{Eugenio Tufino}
\address{Department of Physics and Astronomy, University of Padua, Padua, Italy}
\ead{eugenio.tufino@unipd.it}

\begin{abstract}
This paper presents an approach to introduce physics students to the basic concepts of Large Language Models (LLMs) using Python-based activities in Google Colab. The teaching strategy integrates active learning strategies and combines theoretical ideas with practical, physics-related examples. Students engage with key technical concepts, such as word embeddings, through hands-on exploration of the Word2Vec neural network and GPT-2 - an LLM that gained a lot of attention in 2019 for its ability to generate coherent and plausible text from simple prompts.

The activities highlight how words acquire meaning and how LLMs predict subsequent tokens by simulating simplified scenarios related to physics. By focusing on Word2Vec and GPT-2, the exercises illustrate fundamental principles underlying modern LLMs, such as semantic representation and contextual prediction. 
Through interactive experimenting in Google Colab, students observe the relationship between model parameters (such as temperature) in GPT-2 and output behaviour, understand scaling laws relating data quantity to model performance, and gain practical insights into the predictive capabilities of LLMs. This approach allows students to begin to understand how these systems work by linking them to physics concepts - systems that will shape their academic studies, professional careers and roles in society.
\end{abstract}

\section{Introduction}
Machine learning (ML) and natural language processing (NLP) are increasingly demonstrating their potential in physics education research (PER).
Recent studies, such as those by Fussell et al.~\cite{Fussell2024} proposed a methodology to build confidence in the use of machine coding for analyzing text data, specifically from open-ended survey responses. In a related study, Fussell et al.~\cite{Fussell2024_LabNotes} compared the performance of Large Learning Models, in analyzing students' lab notes through sentence-level classification. Kortemeyer~\cite{Kortemeyer2023} presented the use of LLMs for grading handwritten calculations. 
In another study, Odden et al.~\cite{Odden2020, Odden2021} employed NLP methods to uncover thematic trends in physics and science education research, identifying thematic shifts over time.
At the same time, numerous studies have explored the potential of LLMs as tutors, their application in teacher training, and their evaluation on various physics tasks~\cite{Avila2024, Steinert2024,Kortemeyer2024}, as reviewed by Polverini and Gregorcic~\cite{Polverini2024}. These studies highlight both the versatility and limitations of LLMs, emphasizing the need for critical evaluation of their outputs.
Polverini and Gregorcic further highlight the educational potential of such tools by exploring how understanding the mechanisms behind Artificial Intelligence (AI) models, including large language models (LLMs) like ChatGPT, can inform and enrich physics teaching and learning.
They argue that exposing students to the underlying principles of these models fosters critical thinking and empowers them to use AI tools effectively and ethically in their learning processes.

Although LLMs are becoming increasingly common - while OpenAI's ChatGPT \cite{OpenAI2024_1} is the most studied and widely known to the public, other commercial major LLMs have emerged like Anthropic (Claude) \cite{Anthropic2024}, Google (Gemini)~\cite{DeepMind-Google2024}, and Meta (LLaMA)~\cite{Meta2024} - a clear understanding of how they work is not yet widespread among non-specialists. This knowledge gap underscores the importance of introducing foundational concepts of LLMs in an accessible and pedagogically effective manner.

Building on these lessons, our work shifts the focus from research to instructional applications, aiming to help students understand how AI and NLP models like Word2Vec and GPT-2 operate. Through practical activities, we bridge computational methods and physics concepts, fostering a hands-on understanding of AI’s capabilities and applications.

In this work, we propose an applied approach to help students, even those without specific knowledge of machine learning, understand how deep learning works, particularly in the fields of NLP and Generative AI. The focus is on providing hands-on activities, making these concepts intuitive and engaging.

The activities are structured around Python-based Jupyter Notebooks (JNs) in the Google Colab.
Python was chosen because it is one of the most widely used programming languages, known for its relative simplicity and extensive libraries, which allow students to explore both the coding and conceptual aspects of AI.
Additionally, using Google Colab allows the smooth implementation of this interactive approach, as demonstrated in our previous works~\cite{Tufino2024a,Tufino2024b}.

In fact, as highlighted by Trout and Winterbottom~\cite{Trout2025}, tools like ChatGPT can effectively support students in writing, debugging, and understanding Python code. Similarly, Yeaton et al.~\cite{Yeadon2024} evaluated the performance of GPT-3.5 and GPT-4 on university-level physics coding assignments, demonstrating their potential for solving tasks while highlighting important limitations.

These activities, lasting three hours, were proposed as part of a  module on AI in a Master's degree course on Teaching and Learning Physics. Approximately twenty students participated to the activities. They were from MSc Physics, Physics of Data, or Astrophysics \& Cosmology programs at the university of Padua. In addition, the JNs used in this study are available on GitHub in the folder here \url{https://github.com/etufino/Introduction-to-LLM}. An introductory notebook is also available for students with limited knowledge of Python and JNs. The activities were divided into three main parts:

\begin{enumerate}
    \item Exploring Word2Vec, a neural network model for word embeddings, introduced in 2013 by a team of researchers at Google, which converts words into numerical vectors. This foundational concept in LLMs is critical for understanding how relationships between words—and by extension, concepts—are captured and analyzed. In the context of Natural Language Processing (NLP), Word2Vec exemplifies how semantic relationships can be quantified and visualized, providing a practical entry point into the study of LLMs. Furthermore, this activity highlights the relationships between physics concepts, an important topic in physics education.
    \item Working with GPT-2~\cite{Radford2019}, the transformer-based model by OpenAI that triggered the current revolution in natural language processing. Students will observe GPT-2's context-based predictive nature and interactively modify its parameters to see their effects.
    \item In the third activity, students briefly used the LEAP platform~\cite{Avila2024} to work on physics experiments, using the gpt4o-mini model as a tutor in an experimental activity we proposed. This activity is not discussed further in this article, as it serves as a preliminary exploration.

\end{enumerate}
These activities are designed to be used in the classroom with minimal computing resources and without the need for paid software or advanced hardware, ensuring accessibility for all students. They follow the principles of active learning methods, starting with simple examples and progressively increasing in complexity to encourage student confidence. During the sessions, students worked in groups of two or three. In addition, the exercises include physics-related tasks designed to engage students in meaningful and domain-specific applications. These tasks not only demonstrate the power of AI tools such as Word2Vec and GPT, but also provide opportunities to explore the connection with core physics concepts.
Importantly, they highlight how modern LLMs exhibit emergent properties, such as improved text comprehension, at larger scales.

\section{Word2Vec: From words to numerical embeddings}
One of the key aspects of NLP and LLMs is the concept of embedding, which refers to converting words into numerical representations. Among the most widely used and influential methods for this purpose is Word2Vec, a neural network model designed to learn word embeddings, converting words into numerical vectors that capture their semantic relationships~\cite{WikipediaWord2Vec, Mikolov2013_1}.

The Jupyter Notebook (available on GitHub) uses the Gensim library~\cite{gensim2023} to handle Word2Vec models efficiently and provides the option to choose among three pretrained models: GloVe-Twitter-25, a lightweight model with 25-dimensional vectors; GloVe-Wikipedia-Gigaword-100, a mid-sized model with 100-dimensional vectors; and Word2Vec-Google-News-300, a larger model with 300-dimensional vectors offering more accurate embeddings.
Each model varies in size and scope, with the largest being the Google News model. For this model, the vocabulary size reaches approximately 3,000,000 words, and the vector dimensionality is 300. These characteristics make it particularly rich for semantic exploration, though they may pose challenges in bandwidth-constrained classroom environments.

One of the key functions provided by Word2Vec is \texttt{model.most\_similar}.  This function returns numerical values ranging from 0 to 1, where numbers closer to 1 indicate a higher degree of similarity, suggesting that the concepts represented by the words are closely related. For example:

\begin{verbatim}
model.most_similar("physics", topn=5)
\end{verbatim}

This function returns the top 5 words that are most semantically similar to "physics," providing insights into how the model captures relationships between words. By experimenting with different words and parameters, students can explore the underlying structure of word embeddings and understand how semantic proximity is encoded numerically. This interactive exploration is particularly valuable for grasping how LLMs process and relate textual data.
The numerical values returned by \texttt{model.most\_similar()} represent the cosine similarity between the vector of the input word and the vectors of the most similar words. 
Cosine similarity is a widely used metric in NLP, to measure the similarity between the vector of the input word and the vectors of the most similar words. This concept is familiar to students who have studied vector spaces and linear algebra.

In fact cosine similarity measures the cosine of the angle between two non-zero vectors in a multi-dimensional space. The value ranges from 1 for vectors pointing in the same direction, through 0 for orthogonal vectors, to -1. 
A value close to 1 indicates a high degree of similarity, while values closer to 0 or negative indicate weak or inverse relationships. For example, a similarity score of 0.85 between the words "physics" and "science" reflects their strong connection in the embedding space, whereas semantically unrelated concepts, such as "physics" and "banana," are likely to exhibit values closer to 0 due to the larger angle between their vectors.
\[
\text{cosine similarity}(\mathbf{u}, \mathbf{v}) = \frac{\mathbf{u} \cdot \mathbf{v}}{\|\mathbf{u}\| \|\mathbf{v}\|} = \frac{\sum_{i=1}^{n} u_i v_i}{\sqrt{\sum_{i=1}^{n} u_i^2} \cdot \sqrt{\sum_{i=1}^{n} v_i^2}}
\]

\begin{figure}[h!]
    \centering
    \includegraphics[width=1.0\textwidth]{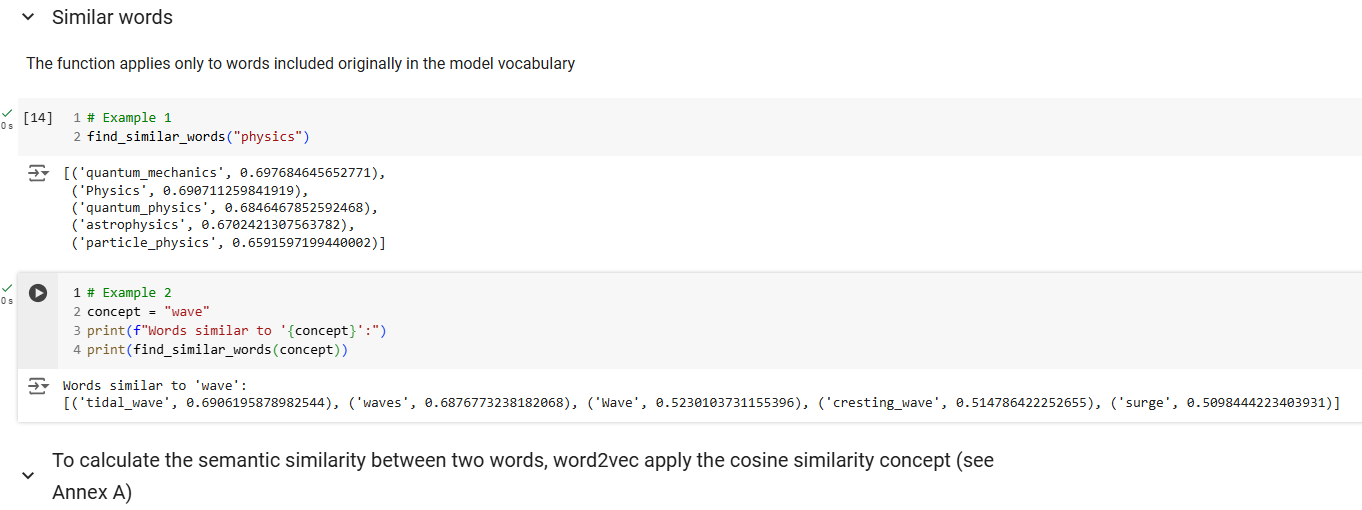} 
    \caption{Examples of using our custom-defined \texttt{find\_similar\_words(word)} function in Word2Vec with the words “physics” and “wave,” showing the top five results.}
    \label{fig:similarword}
\end{figure}

In the Jupyter notebook, we introduced some auxiliary functions (with the model and the topn parameter already set). For instance, an instructive function is \texttt{analogy(word1, word2, word3, topn=1)} - internally calling the Word2Vec built-in \texttt{model.most\_similar()} method - which finds words that complete analogies based on vector relationships.
This function allows us to solve analogies of the type "word1 is to word2 as word3 is to ?". Essentially, we are using the fact that the analogy can be expressed as a vector operation (see section \ref{sec:vector_operation}).
\[
\mathbf{word2} + \mathbf{word3} - \mathbf{word1}
\]

and returns the word whose vector is closest to this resulting vector.

\begin{figure}[h!]
    \centering
    \includegraphics[width=1.0\textwidth]{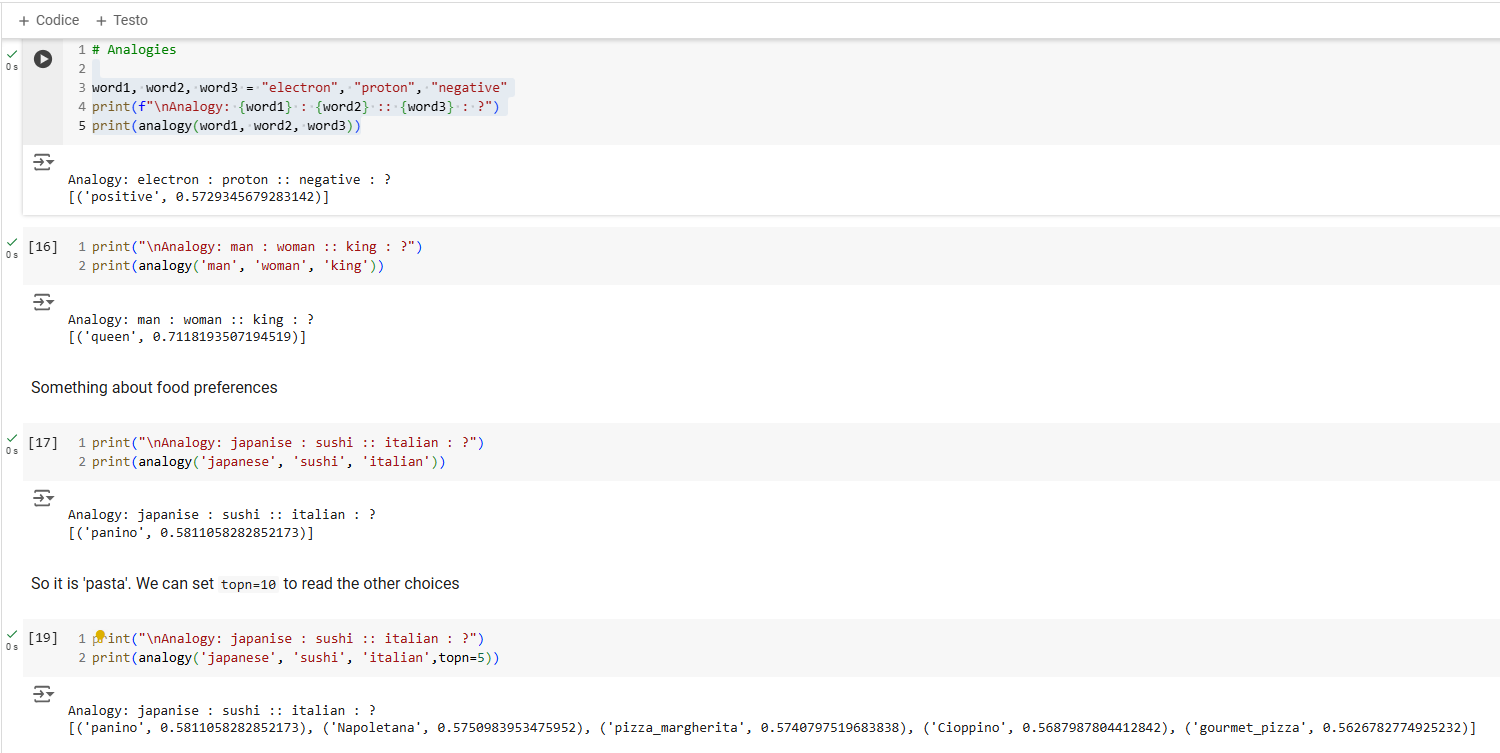} 
    \caption{Examples of analogies from the JN using the custom-defined function \texttt{analogy(word1, word2, word3, topn=1)}, both in physics-related contexts and in everyday scenarios. }
    \label{fig2:example_analogy}
\end{figure}

For example, Figure \ref{fig2:example_analogy} explores the analogy: "electron is to proton as negative is to ?", providing insights into the relationships captured by word embeddings. Students can use this function to experiment with analogies, gaining a deeper understanding of how semantic relationships are represented numerically.

As reported in Mikolov et al.~\cite{Mikolov2013_1,Mikolov2013_2}), the researchers observed that word vectors created by their network successfully captured the analogy between a country and its capital. For example, the relationship France: Paris = Italy: Rome emerged naturally, despite the model not being explicitly provided with the concept of a "capital city." This relationship was inferred during the training phase of the network, based on patterns in a sufficiently large corpus of data.

\subsection{Biases from data}

For instance, running the following command:
\begin{verbatim}
print("\nAnalogy: man : surgeon :: woman : ?")
print(analogy('man', 'surgeon', 'woman'))
\end{verbatim}
Returns \texttt{ [('nurse', 0.64670729637146)]}.
This outcome highlights how the model captures relationships from its training data, but it also reflects  cultural or contextual relationships and inherent biases. The association of "woman" with "nurse" rather than "surgeon" reveals societal stereotypes encoded in the training corpus. This issue is not limited to Word2Vec; it extends to larger and more advanced LLMs, such as GPT-4 and similar models, which often inherit biases from the massive datasets they are trained on. These biases can affect their outputs in subtle yet impactful ways, influencing how they generate responses and complete tasks.

Understanding and addressing training data bias is a critical challenge in the development and deployment of LLMs. It requires not only curating diverse and representative datasets but also implementing debiasing techniques within the models themselves. For educators, exploring these biases provides an opportunity to discuss the ethical and technical challenges associated with LLMs and to emphasize the importance of critical engagement with AI tools.
Mistakes in analogies can often be attributed to limitations in the training dataset or the model itself. For instance, insufficient emphasis on semantic relationships in the corpus, ambiguity from words with multiple meanings, or the static nature of Word2Vec embeddings—where each word has a single vector regardless of context—can all contribute to errors. These limitations, along with examples and explanations, are further explored in the related JN.
\subsection{Visualize word vectors in a 2D space}
The word vectors generated by Word2Vec exist in a high-dimensional space. In the largest model used by students in  the JN, the vector size is 300. To visualize these vectors, the JN uses a reduction to 2D using Principal Component Analysis (PCA) technique. While PCA effectively reduces the 300-dimensional space to 2D for visualization, it may lose important relationships between words in the process. Thus, PCA is more suited for qualitative insights rather than precise analysis.
To explore semantic relationships, we grouped words into four major physics topics and visualized them in 2D. The groups included:

\begin{verbatim}
mechanics = ['force', 'mass', 'acceleration', 'velocity', 'momentum']
thermodynamics = ['heat', 'temperature', 'energy', 'pressure', 'work']
electromagnetism = ['charge', 'current', 'voltage', 'magnetic', 'electric']
quantum = ['quantum', 'wave', 'particle', 'uncertainty', 'atom', 'photon']
\end{verbatim}

\begin{figure}[h!]
    \centering
    \includegraphics[width=0.9\textwidth]{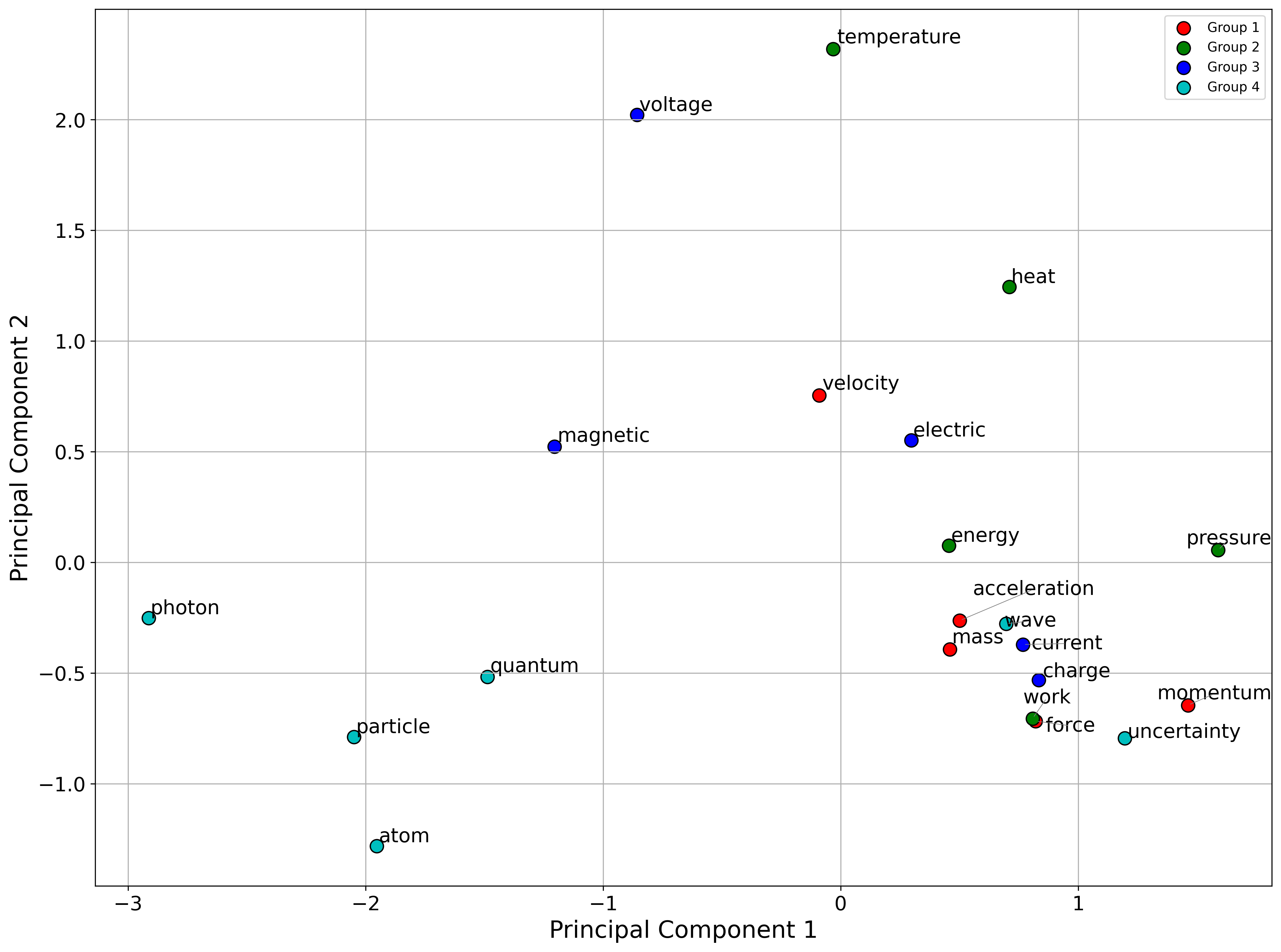} 
    \caption{2D Visualization of Word Embeddings. Words that appear closer in the plot are semantically closer in the embedding space. Notice how the “quantum” group forms a distinct cluster, whereas “work” and “force” nearly overlap, indicating they are represented as very similar vectors in this model.}
    \label{fig:2D visualization groups}
\end{figure}

The visualization of word embeddings reveals interesting relationships between the groups of words (Figure \ref{fig:2D visualization groups}). Notably, the quantum group is well-separated from other clusters, reflecting the unique semantic nature of quantum-related terms. However, wave and uncertainty show connections to terms in other groups, likely due to their interdisciplinary usage across mechanics and electromagnetism.

Another interesting observation is the near overlap of work and force, emphasizing their strong semantic similarity in the embedding space. Similarly, the words within the mechanics group, such as "force," "momentum," and "velocity," are tightly clustered, reflecting their close conceptual relationships in physics.

These patterns illustrate how Word2Vec captures semantic proximities, while also highlighting potential limitations when similar terms (e.g., "work" and "force") become indistinguishable in lower-dimensional spaces.


\subsection{Vector operations in the semantic space}
\label{sec:vector_operation}
In Word2Vec, it is possible to perform vector operations in the semantic space, which is closely related to the analogies discussed earlier. For example, the well-known analogy \textit{king:man = ?:woman} illustrates how word embeddings capture relationships between words. This analogy shows that the difference vector between "king" and "man" is approximately aligned with the difference vector between "queen" and "woman": \[
\mathbf{king} - \mathbf{man} + \mathbf{woman} \approx \mathbf{queen}
\]

reflecting the semantic relationship encoded in the embedding space. 
Additionally, the vector connecting "king" and "man" is approximately parallel to the vector connecting "queen" and "woman", further highlighting the consistency of these relationships in the embedding space (see the related Figure in the Jupyter notebook).
As an exercise, students can also compute the cosine similarity between "Result" and "queen" to quantify their closeness.
Considering the previous example regarding countries and capital, the difference vector between “Paris” and “France” closely matches the difference vector between “Rome” and “Italy”.

It is important to emphasize that in Word2Vec, the relationships between vectors reflect semantic associations rather than quantitative ones. This allows for the exploration of how physical concepts are connected within natural language, rather than representing equations or physical laws.

At the end of the exploration of Word2Vec, we discussed with the students the concept of static embeddings, highlighting how each word is represented by a single fixed vector, regardless of the context. For example, the word \textit{mole} can have different meanings: a small animal (biology), a unit of measurement (chemistry), and a mark on the skin (medicine).
The students noted that Word2Vec assigns the same embedding to "mole" in all these contexts, revealing a limitation of the model. This reflection prepared the students to understand more advanced models, such as GPT-2, which utilizes dynamic embeddings that adjust to the context.

\subsection{Training a Word2Vec model from scratch}
A common question asked by students is how word vectors are constructed. It is possible to explore the components of any vector in the semantic space using a simple command, such as inspecting the vector for "physics" with model['physics']. This provides a numerical representation of the word in the embedding space, illustrating the well-known statement by John R. Firth: "You shall know a word by the company it keeps"~\cite{Firth1957}. Word2Vec exploits this idea by using the context in which words appear to capture their semantic relationships and encode them as vectors.
A more formal approach to understanding word similarity involves examining conditional probabilities of co-occurrence. Specifically, if $P(a \mid c)$ is defined as the probability that word a appears in the vicinity of word  c, then a and b can be considered semantically similar if
$P(a \mid c) \approx P(b \mid c) \text{ for every } c$
By modeling how frequently words appear in the same contexts, Word2Vec and similar embedding methods effectively learn these conditional probabilities, thus capturing semantic relationships in a low-dimensional vector space.

To deepen students' understanding, it is possible to propose a hands-on supplementary activity (Part 2 in the JN). 

Word2Vec is a neural network, specifically a shallow, two-layer neural network~\cite{Jurafsky2024}. Its weights and biases are initially random, and training is required to optimize them.
Students can build a Word2Vec model from scratch by providing the neural network with a corpus of data, which can be as small as a few pages or a chapter from a book. Starting from random initial values, the model gradually adjusts the vectors for each word during the training process by minimizing a cost function. This function ensures that the embeddings capture meaningful relationships between words as they co-occur in the corpus.

The two primary training techniques used in Word2Vec are CBOW (Continuous Bag of Words) and Skip-Gram, which define how the model predicts word contexts or target words during training~\cite{Mikolov2013_1}. CBOW predicts a target word based on its surrounding context, while Skip-Gram predicts the context given a target word.
The code of the Jupyter Notebook  allows students to apply them to an assigned corpus. In general, the data corpus must be prepared through tokenization and text preprocessing—splitting the text into individual words (tokens), removing extraneous symbols, and standardizing the format (e.g., converting to lowercase). This step is essential for ensuring the data is in a suitable format for training a Word2Vec model.
Thanks to common Python NLP libraries, such as NLTK, basic preprocessing can be done. As a first exercise for demonstrating the training procedure of a Word2Vec network, we created a corpus of sentences, taken from a Wikipedia page. These sentences were preprocessed. 
Since our custom quantum mechanics corpus of sentences is very small, the results from the \texttt{model.most\_similar} method are not especially meaningful. Both CBOW and Skip-Gram yield very similar outputs. Nevertheless, the exercise effectively demonstrates how to train a Word2Vec model and compare the two architectures in practice.

As a second exercise, students can use the Text8 corpus to demonstrate how a larger and more varied dataset can significantly improve both the quality and the interpretability of the learned embeddings. Text8 is an excerpt from the English Wikipedia, cleaned of punctuation and already tokenised. Its size and structure make it particularly suitable for showing how both CBOW and Skip-Gram behave on a larger corpus.
For example, when comparing similarities for words like "physics" using CBOW vs. Skip-Gram on text8, students can see more domain-relevant neighbours than in the smaller custom corpus.
\noindent
\textbf{Listing \ref{lst:physics-similar}} shows the top 5 most similar words to “physics” with both CBOW and Skip-Gram. 
We can see that both architectures capture semantically related terms, indicating that 
the embeddings successfully learn relevant context for “physics.”

\begin{lstlisting}[language=Python, caption={Top 5 most similar words to “physics” using CBOW and Skip-Gram on the text8 corpus.}, label={lst:physics-similar}]
Words similar to 'Physics' (CBOW): [('mechanics', 0.79958), ('chemistry', 0.77116), ('mathematics', 0.74596), ('astronomy', 0.74021), ('theoretical', 0.73904)]
Words similar to 'Physics' (Skip-Gram): [('electrodynamics', 0.82742), ('electromagnetism', 0.81008), ('mechanics', 0.79786), ('chemistry', 0.79652), ('supersymmetry', 0.76048)]
\end{lstlisting}

For comparison, searching for similar words to "pulsar" - a rarer term - shows that Skip-Gram tends to capture relationships better for low-frequency words. This practical exercise further demonstrates how corpus size, word frequency and training architecture can influence the semantic quality of the resulting embeddings.
As shown in Listing~\ref{lst:pulsar-similar}, Skip-Gram provides more domain-specific neighbors for the relatively rare term \textit{pulsar}, demonstrating its effectiveness for low-frequency words.

\begin{lstlisting}[language=Python, 
                   caption={Top 5 words most similar to "pulsar", comparing CBOW vs. Skip-Gram embeddings on the text8 corpus.}, 
                   label={lst:pulsar-similar}]
Words similar to 'pulsar' (CBOW): [('pchar', 0.803925096988678), ('aarp', 0.7837557792663574), ('pdes', 0.7755052447319031), ('cauer', 0.7689059376716614), ('calorimetric', 0.7668492794036865)]

Words similar to 'pulsar' (Skip-Gram): [('pulsars', 0.8908295035362244), ('extrasolar', 0.8871657252311707), ('ephemeris', 0.8636508584022522), ('occultations', 0.8625813722610474), ('lensing', 0.8560528755187988)]

\end{lstlisting}

Students were able to check this also by experimenting with the three pre-trained Word2Vec models suggested in the Jupyter notebook (Twitter model, Wikipedia model and Google News model). These examples show that network trained on large datasets, such as the Google News model which is approximately 1663 MB in size, tend to produce more accurate word embeddings. 
This activity illustrates how Word2Vec generates semantic representations from textual data and the importance of factors such as corpus size and domain relevance.

\section{Exploring GPT-2: Text Generation and stochastic predictions}
In the next part of the module, students worked with the second provided Jupyter notebook, focusing on GPT-2, the generative model released by OpenAI in 2019. 
Although the GPT-2 is a lot simpler than the actual more advanced models such as the GPT-4, it is based on the same Generative Pre-trained Transformer (GPT) architecture. This makes it an ideal choice for educational purposes, as it is computationally manageable on personal computers without requiring advanced hardware or network resources.
GPT-2 works as a stochastic predictive model, generating text word by word by predicting the most likely next word based on the given context. This probabilistic nature allows the model to produce different outputs for the same prompt, depending on the random sampling process. By adjusting some parameters in the Jupyter Notebook students were able to explore how these factors influence the consistency and quality of the generated text. 
The Transformer architecture, which relies on self-attention mechanisms to capture long-range dependencies in text, was only briefly mentioned to the students~\cite{Vaswani2017}.
In the Jupyter Notebook, we use specific Python libraries to interact with GPT-2, starting with the Hugging Face "Transformers" library.
Within the Notebook, three variants of GPT-2 are discussed:
gpt2-small (the smallest version), gpt2-medium, gpt2-large.
Gpt2-large is significantly larger than the other two variants, but it is still feasible to use in the classroom on a standard personal computer and via Google Colab.

GPT-2 Large contains approximately 774 million trainable parameters (they are the adjustable elements within a neural network that the model learns and updates during the training phase) with an embedding vector dimension of 1280 and vocabulary size of 50257 token.

The Hugging Face library provides not only pre-trained models, but also tokenisers that convert input text into sequences of tokens (i.e. numerically encoded pieces of text). 
To further illustrate the concept of tokens, students were tasked to tokenize the phrase "What is AI?". This sentence, which breaks down into four tokens (What, is, AI, ?), was also visualized using the OpenAI Tokenizer page, allowing students to explore how text is pre-processed before being fed into the model.

\subsection{GPT-2 Activity: Paragraph Generation}
\begin{figure}[h!]
    \centering
    \includegraphics[width=0.9\textwidth]{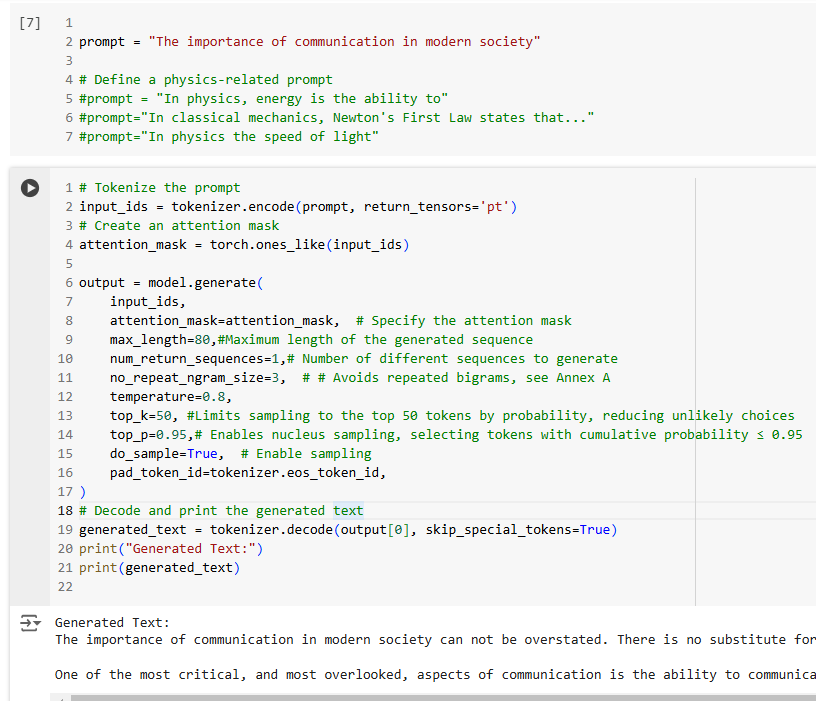} 
    \caption{An example of a GPT-2 code cell that generates a paragraph for the prompt: "The importance of communication in modern society."}
    \label{fig:gptparagraph}
\end{figure}

In the first exercise (Figure~\ref{fig:gptparagraph}), students used GPT-2 to generate a paragraph from the generic prompt \textbf{"The importance of communication in modern society"}. An example of generated output is shown below:
\begin{quote}
The importance of communication in modern society cannot be overstated. There is no substitute for being able to communicate effectively with others. In this way, we can gain insight, learn from others, and gain better insight into ourselves. One of the most critical, and most overlooked, aspects of communication is the ability to communicate emotion. Emotion is a powerful tool in understanding human behavior...
\end{quote}

When the same cell was run several times, students noticed that the text generated varied significantly, demonstrating the stochastic nature of GPT-2's output. Students then tried more specific physics prompts, such as "\textbf{In physics, Newton's laws describe the relationship between motion and forces}" or "\textbf{The theory of relativity, proposed by Albert Einstein, revolutionized our understanding of space and time}".
In these cases, the results were less reliable and very often provided information that was incomplete, meaningless or inaccurate.
For instance, when using the prompt on the theory of relativity, GPT-2 generated the following output instance:
\begin{quote}
The theory of relativity, proposed by Albert Einstein, revolutionized our understanding of space and time. The laws of gravity, which govern the motion of the Earth around the Sun, are not the only thing that were changed by this theory. The speed of light, the speed at which an object can move in one instant, changed from 186,000 miles per second to 299,792,458 miles per...
\end{quote}
This response contains a number of inaccuracies, including the assertion that the speed of light 'changed' as a result of the theory of relativity, and a confusing mix of units (miles per second and metres per second). These errors demonstrate GPT-2's tendency to generate text that, although plausible, is factually incorrect or misleading when faced with technical or domain-specific prompts. They were then asked to reflect on questions such as:

\textit{How coherent is the generated text with the given prompt?
Does the model really understand the concepts, or is it just generating plausible sentences?}

Students were encouraged to experiment with the model by adjusting selected parameters in the notebook. In order to focus and simplify the exploration, only the following parameters were asked to vary:
\begin{itemize}
\item "Temperature"\footnote{It's important to note that the "temperature" parameter in language models like GPT-2 is not directly related to physical temperature, but 
it is a technical term borrowed from statistical mechanics.}: Higher temperatures produce more random (and creative) outputs, while lower temperatures yield more predictable results. 
\item Max Length: Adjusts the length of the generated text.
\item Top-k:  Limits the selection to the top-k most probable tokens, reducing randomness while maintaining diversity. For example, $top\_k=50$ restricts choices to the top 50 tokens ranked by probability at each step.
\end{itemize}
They were asked to notice the most meaningful text generated for their specific physics context, reflecting on how the parameter influenced the behavior of the model.

\subsection{Visualizing token probabilities and temperature Effects in GPT-2}
\begin{figure}[htbp]
    \centering
    \begin{minipage}{0.45\textwidth}
        \centering
        \includegraphics[width=\textwidth]{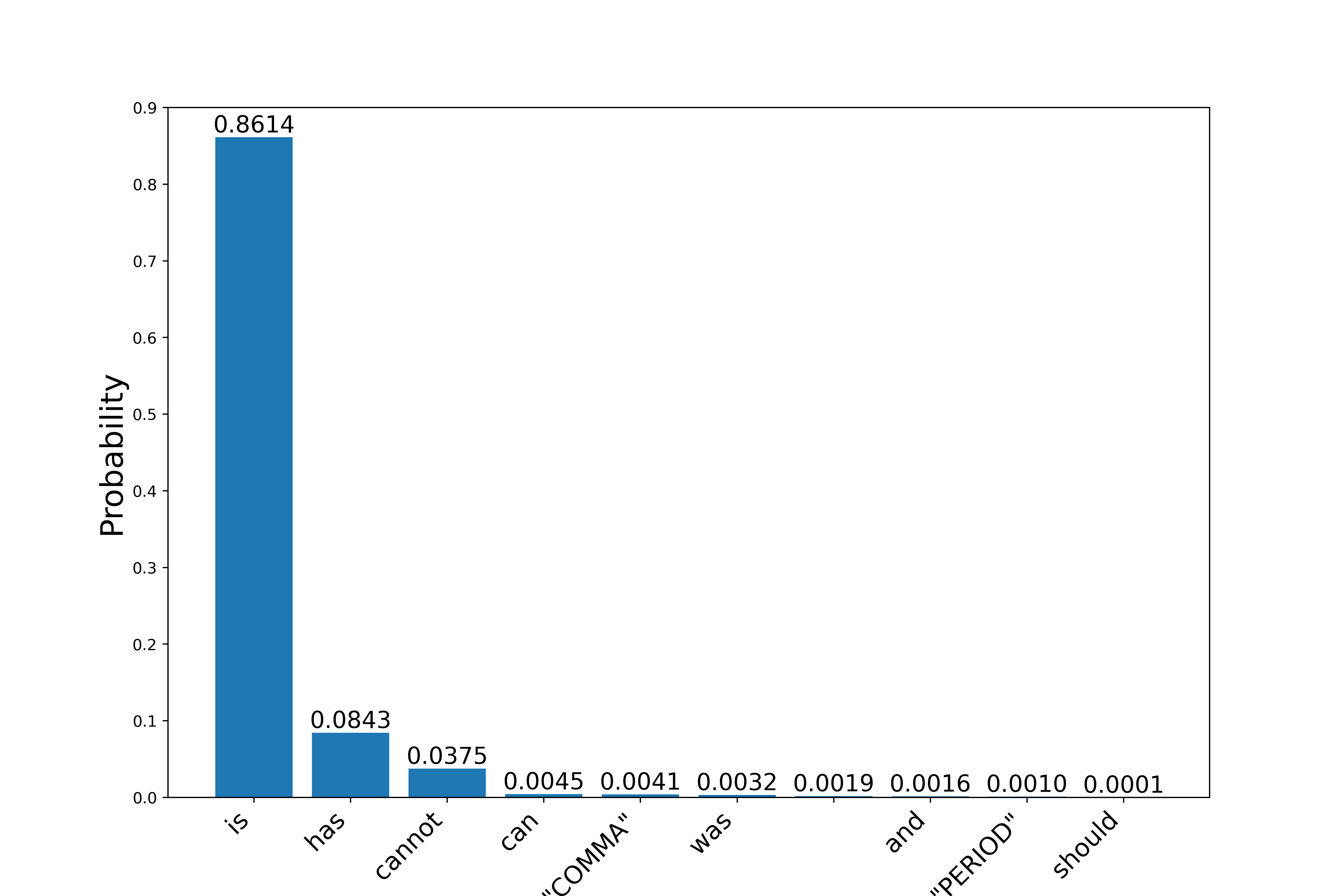}
        \caption{Probability distribution of the top 10 most likely next tokens with temperature = 0.4. Lower temperature results in a more concentrated probability distribution.}
        \label{fig:figuraHist0.4}
    \end{minipage}
    \hfill
    \begin{minipage}{0.45\textwidth}
        \centering
        \includegraphics[width=\textwidth]{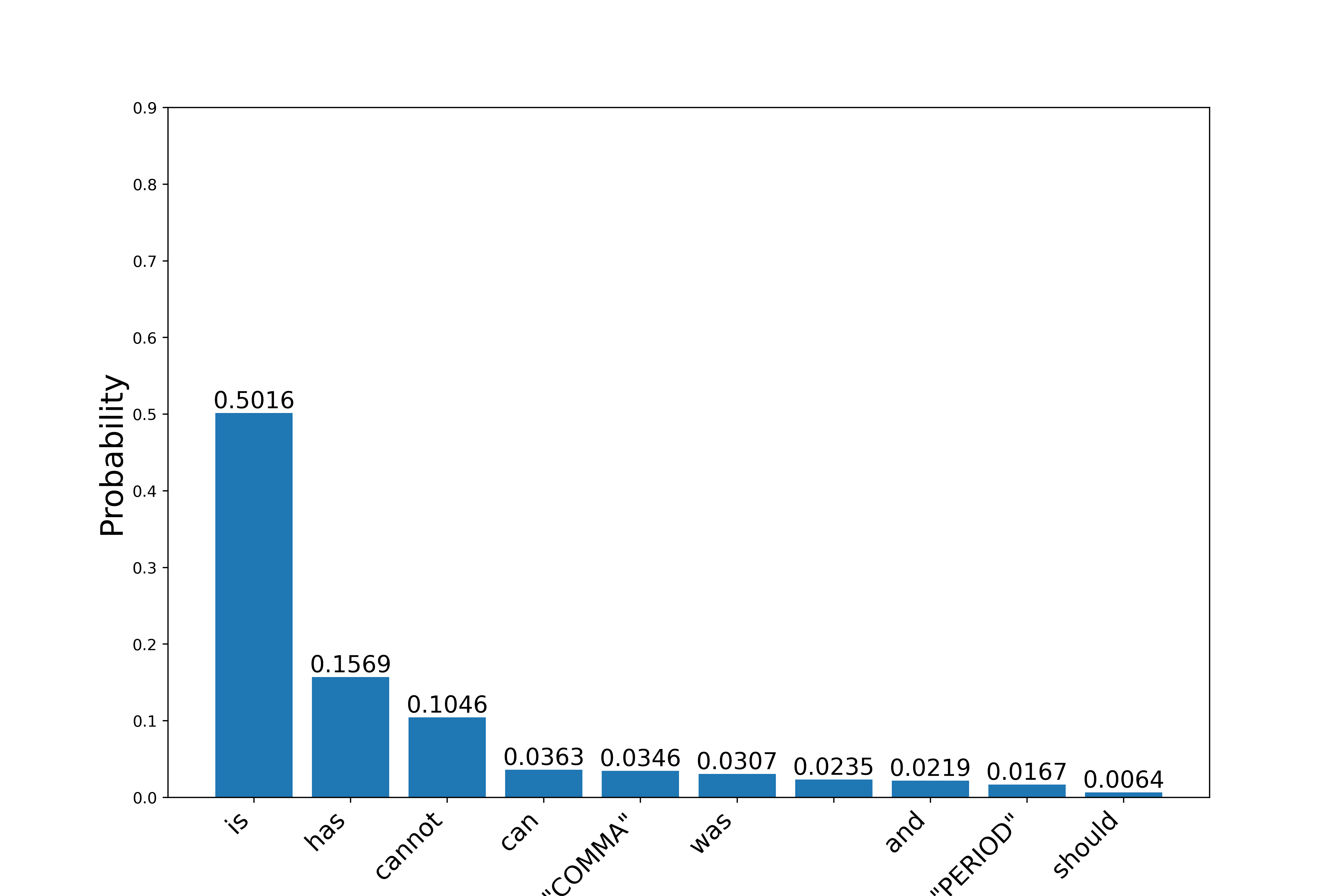}
        \caption{Probability distribution of the top 10 most likely next tokens for the same prompt with temperature = 0.8. Higher temperature leads to a more spread out probability distribution.}
        \label{fig:figuraHist0.8}
    \end{minipage}
\end{figure}

In the next exercise, students explored  how GPT-2 assigns probabilities to potential next words and how these probabilities are influenced by the temperature parameter.
For example, students worked with prompts such as "The importance of communication in modern society" or one of the physics-specific prompt. The notebook allowed them to visualize the probabilities for the next token, with an option to display the top 30 most probable tokens (if top\_k=30) for a given prompt.
By experimenting with the temperature parameter, students observed how it affects the distribution of probabilities:
\begin{itemize}
\item Lower temperatures result in more predictable and focused outputs, concentrating probabilities on fewer tokens.
\item Higher temperatures increase randomness, spreading probabilities across more tokens and enabling more creative, but potentially less coherent, outputs.
\end{itemize}
To visualize this effect, students could plot histograms showing the top 10 most probable tokens and their respective probabilities. 
Figures \ref{fig:figuraHist0.4} and \ref{fig:figuraHist0.8} illustrate the differences in token probability distribution at temperature = 0.4 and temperature = 0.8, respectively for the prompt: "The importance of communication in modern society".

In another exercise, students can generate progressively longer paragraphs by iteratively sampling tokens using a code snippet included in the Jupyter Notebook (omitted here for brevity). Building upon their earlier work with temperature, they can explore additional parameters—such as \texttt{no\_repeat\_ngram\_size} listed in the Notebook’s Appendix. 

To conclude this activity with GPT-2, students explored the importance of model size by experimenting with the three available GPT-2 variants (small, medium and large) and found that larger models tended to produce more coherent and contextually appropriate text. 
In this way, students can observe in a simplified context the relationship between the amount of data used to train LLMs and their performance, a relationship that is also observed in larger and more recent models, known as scaling law~\cite{Kaplan2020}.
It also showed that generating meaningful text for physics-related prompts is more challenging than for general prompts. Readers can also compare this practical experience with OpenAI's GPT-3 Playground environment, which uses GPT-3 and allows users to adjust parameters such as temperature~\cite{Polverini2024}. However, the Playground environment does not offer the same level of configurability and interactivity as the Jupyter Notebook environment and requires registration on the OpenAI platform.

\section{Conclusions}
This work proposes an approach for familiarizing students with the foundational concepts of Large Language Models (LLMs) using Python-based interactive activities. Through hands-on exercises with Word2Vec and GPT-2, students developed a practical understanding of LLM's word embeddings and the statistical mechanism of next-token prediction inherent in GPT-2. These activities provided a concrete experience of how LLMs process and generate text, highlighting the role of various parameters such as temperature and model size in influencing output coherence and creativity.
Since the release of GPT-2 in 2019, which marked a significant advance in large language models (LLMs), it has received widespread surprise and attention from non-technical media. OpenAI's blog post~\cite{OpenAI2019} detailed the model's capabilities and the considerations that influenced its staged release, including concerns about its potential misuse for generating fake news.
In the following years there have been remarkable improvements in LLMs. The field is evolving rapidly, requiring continuous updates and critical reflection in educational interventions. For example, ongoing research efforts aim to overcome the limitations associated with using tokens as input to the Transformer architecture. 
The latest OpenAI model, part of the new ``o'' series released in September 2024~\cite{OpenAI2024}, while still based on the Transformer architecture and the token prediction mechanisms of their predecessors that we have illustrated, also incorporate ``thinking'' phases, enhancing their ability to perform complex tasks and engage in more sophisticated interactions.
Similarly, Google has introduced its own reasoning-focused AI mode, called Gemini 2.0 Flash Thinking Experimental, incorporating mechanisms that allow it to fact-check itself, thereby reducing the occurrence of repetitive errors~\cite{Google2024}. 

The interactive Python activities designed for this study were effective in engaging students and fostering a deeper appreciation of the technologies.
Students were able to experiment with different size GPT-2 models observing firsthand how model size affects the generation of coherent and contextually appropriate text. 
In addition, these activities can be adapted for undergraduate or high school students. Future work could focus on evaluating the impact of these interactive exercises and exploring ways to extend them. By equipping students with a basic understanding of Large Language Models, we prepare them to use these models effectively in their research projects and professional careers.

\section*{Acknowledgements}
I would like to sincerely thank Dr Marta Carli for giving me the opportunity to pilot test these activities in December 2024 within her MSc course "Teaching and Learning Physics" at the University of Padova.

\end{document}